\documentclass[journal,twoside,web]{ieeecolor}
\usepackage{generic}
\usepackage{cite}
\usepackage{amsmath,amssymb,amsfonts}
\usepackage{algorithm}
\usepackage{algorithmic}
\usepackage{graphicx}
\usepackage{textcomp}
\usepackage{multirow}
\usepackage[table,xcdraw,dvipsnames,svgnames,x11names]{xcolor}
\usepackage[final]{changes}

\definechangesauthor[color=Green]{E}
\usepackage{todonotes}
\setcommentmarkup{\todo[color={authorcolor!20},size=\scriptsize]{#3: #1}}

\def\BibTeX{{\rm B\kern-.05em{\sc i\kern-.025em b}\kern-.08em
    T\kern-.1667em\lower.7ex\hbox{E}\kern-.125emX}}
\begin{document}
\title{Application of blockchain for secure data transmission in distributed state estimation}
\author{S. Asefi, Y. Madhwal, Y. Yanovich, and E. Gryazina
\thanks{S. Asefi and E. Gryazina are with center for energy science and technology (CEST), Skolkovo institute of science and technology (Skoltech), Moscow, Russia (e-mail: sajjad.asefi@skoltech.ru), (e-mail: e.gryazina@skoltech.ru). }
\thanks{Y. Madhwal and Y. Yanovich are with center of computational and data science and engineering (CDISE), Skolkovo institute of science and technology (Skoltech), Moscow, Russia (e-mail: yash.madhwal@skoltech.ru), (e-mail: y.yanovich@skoltech.ru). }}
\maketitle

\begin{abstract}
\replaced[]{The application of renewable energy sources in the power grid increases the necessity of tracking the system's state, especially in smart grids, where there is a bidirectional transfer of data and power.}{The application of renewable energy sources in the power grid, especially in smart grids, where there is a bidirectional transfer of data and power, increases the necessity of tracking the system's state.} \deleted[]{Additionally,}\replaced[]{The}{the} complexity of coupling between communication and the electrical infrastructure in a smart grid \replaced[]{will create a higher chance for security breach}{will allow a vast amount of vulnerabilities}. Increasing the state estimation accuracy will help the smart grid operator efficiently manage the system. \replaced[]{The paper proposes}{In this paper, we propose} an \replaced[]{integration}{combination} of distributed state estimation \replaced[]{with}{and} a blockchain designed communication platform. Additionally, the asynchronous manner for data transmission, which is more likely to happen in the real world, has been considered \replaced[]{as the second task of}{the second case in} this \replaced[]{research}{work}. Finally, a detailed analysis \replaced[]{of the blockchain-based application}{regarding the application of blockchain} in distributed state estimation \replaced[]{is}{has been} provided. The numerical analysis shows that the proposed method meets real-world performance requirements and brings high security and reliability to the distributed state estimation process.
\end{abstract}


\section{Introduction}

% Lorem ipsum \added[,comment={lipsum, lipsum}]{dolor sit amet}, consectetur adipiscing elit. Integer luctus molestie hendrerit. Nullam id consequat turpis. Curabitur vitae suscipit ipsum, et lobortis quam. Curabitur convallis scelerisque erat, a pellentesque eros luctus dapibus. \deleted[]{Suspendisse consectetur ligula in rutrum pellentesque.} Cras hendrerit vel nunc non consequat. Curabitur nec est dolor. Cras vulputate commodo augue, sit amet tempus diam posuere a. Cras eget libero placerat, consequat justo non, consequat diam. Fusce nisi turpis, dictum vel eros nec, egestas imperdiet tellus. Duis non ligula velit. Suspendisse nec tellus tellus.\note[]{Proin a lectus vestibulum, mollis eros id, vehicula mauris.}

% Phasellus molestie accumsan rhoncus. Mauris eu turpis non leo rutrum tempus ut sed dolor. Vestibulum ante ipsum primis in faucibus orci luctus et ultrices posuere cubilia Curae; \replaced[]{Donec et massa in arcu sagittis posuere a a erat.}{Vivamus tincidunt nec purus vitae porttitor.} Pellentesque molestie dolor ultrices tincidunt mollis. In hac habitasse platea dictumst. Aliquam molestie, sem eu suscipit pharetra, odio turpis tristique odio, molestie egestas sapien massa in elit.\note[]{Fusce eget sagittis eros. Donec a posuere elit, vitae congue augue.} Donec semper mi felis, id cursus turpis pretium in.

\label{sec:introduction}
\IEEEPARstart{P}{OWER} grid has faced different challenges due to the increasing utilization of the distributed energy sources, and \deleted[]{at the same time,} the non-stop escalating level of energy demand \cite{musleh2019blockchain}. Hence, state estimation (SE) plays an essential role in justifying and regulating system operator decisions like economic dispatch, load frequency control, electricity markets, and load forecasting. \added[]{It is to be noted that advanced SE can improve monitoring and controlling the power grid in case of a contingency. Especially for the smart grids in which bidirectional transfer of electrical energy and system/consumer data \deleted[]{will} increases the complexity} \cite{gomez2018electric, kurt2019secure}. Such a system can be divided into two integrated parts, \replaced[]{i.e.,}{including} physical equipment of a traditional power system \added[]{(Physical part)}, \deleted[]{i.e. physical \replaced[]{part}{system}}, and telecommunication equipment\added[]{ (Cyber part)} \deleted[]{, i.e. cyber \replaced[]{part}{system}}. Combining these two parts will lead to a cyber-physical power system (CPPS) \cite{yohanandhan2020cyber}. \deleted[]{Bidirectional transfer of electrical energy and system/consumers data will increase the complexity and, consequently, the critical CPPS's vulnerability.}
\subsection{Power system state estimation}
Although SE is highly comparable to the conventional load flow, \replaced[]{it}{SE} considers the unpredictable errors that might originate due to unexpected system changes, meter\added[]{s} or communication system errors, inaccuracy in equipment calibration, planned manipulation from a malicious attacker\replaced[]{,}{and} etc. \cite{schweppe1970power, kurt2019secure}. \added[]{\replaced[]{Additionally}{In addition to their conceptual difference}, conventional load flow analysis does not consider redundancy and imprecision of the system's measurement data, whereas SE considers the mentioned features} \cite{schweppe1970power}. \deleted[]{The reliability of estimated states is one of the critical features of a power grid.} \deleted[]{However, monitoring and controlling the power grid can be improved by advance SE in case of a contingency.}

\replaced[]{Considering a brief background of}{In order to provide a brief history on} the evolution of the SE method and its application in power system, \replaced[]{it is worth noting}{we must mention} that as soon as Schweppe pointed out the application of SE in power system, \replaced[]{it attracted industrial communities' attention}{industry instantly accepted it} \cite{wu1990power}. However, they did not apply the basic weighted least squares (WLS) method proposed by him at the beginning but \replaced[]{utilized}{implemented} two \added[]{different} methods, developed by Dopazo \cite{dopazo1970state}, and Larson \cite{larson1970state}. However, after a while revised version of Schweppe’s method was admitted, and those two methods were \replaced[]{excluded}{rejected} from industrial applications \cite{allemong1982fast}. Taking into account the sparsity of the gain matrix, different researchers made several attempts to \added[]{slightly} improve the WLS estimator \deleted[]{slightly} \cite{garcia1979fast}. Also, SE \replaced[]{related issues}{problems} like ill-conditioning appeared afterwards, and several methods such as the inclusion of zero injection power within constraints of SE were proposed to overcome these issues \cite{aschmoneit1977state, gomez2018electric, abur2004power}. \deleted[]{The WLS based SE  }Schweppe et al. proposed applying non-quadratic estimators for bad data detection within the early years \deleted[]{of application}. Later on, there were plenty of studies on methods for identifying \replaced[]{bad}{insufficient} data, such as application of least absolute value \deleted[
%, comment=Maybe delete this
]{for this purpose}\cite{irving1978power}. \added[]{It is to be noted that} from the early stages of utilizing SE, the mentioned areas and problems have been an active research area for \deleted[]{energy management systems (where mainly the SE is done in power system) and in} the power systems\added[]{, operations and control} research community.

\replaced[]{Growth}{The expansion} of the power system due to increase in the level of needed electricity consumption and propagation of the \added[]{communication} technology, \replaced[]{bring in}{there are} several problems assigned with power system operation, especially centralized SE \added[]{(CSE)} such as\deleted[]{,} \textit{Expansion of power system continent-wise}, \textit{Policy and privacy}, \textit{Dimension of the grid}, \textit{Communication bottleneck}, \textit{Data size} and \textit{Security/Reliability}, to name a few.

Expansion of the power grid over continents makes an interconnected system that \replaced[]{these continents can be affected by contingencies in other ones}{these areas can be affected by contingencies in other areas} \cite{conejo2007optimization}. \replaced[]{Although, in some regional expansion cases}{The spread of the network over countries or continents}, e.g. the case with regional transmission organizations (RTOs) in Europe, \added[]{operators} are using HVDC technology for power transfer which is also another research area for considering hybrid HVDC/AC \deleted[]{transmission} SE so that they can meet the characteristics of the new network regarding Supervisory Control and Data Acquisition (SCADA) \added[]{system} \cite{ayiad2020state}. Vulnerability and inflexibility of \replaced[]{CSE}{centralized SE}  \replaced[]{makes it}{make centralized SE} unsuitable for a multi-area (or multi RTO) estimator from \deleted[]{a} policy and privacy point of view \cite{conejo2007optimization}. The grid's high dimension is another challenge that affects the computational difficulty \cite{pau2017efficient, xu2016robust}. Having only one central control unit, extensive network parameters and measurement unit’s information, \replaced[]{which needs to be transferred}{transfer} to this \replaced[]{unit}{central coordinator}, \added[]{may result in} communication bottlenecks \cite{zhang2020robust, minot2015distributed}. Another problem that has attracted the researcher’s attention, especially in smart grids, is that the size \deleted[]{(big data)} and the speed of receiving data \added[]{(so called big data)} from measurement units might \replaced[]{be infeasible to be stored and processed}{not be store and process in a single unit} \cite{kurt2019secure, rostami2017distributed}. Moreover, in most of the literature, it is assumed that the central node is secure, though it can be the most vulnerable, insecure and unreliable point in a network and prone to a single point of failure \cite{kurt2019secure ,kekatos2012distributed}.

\subsection{Power system distributed state estimation}
One way to \replaced[]{overcome CSE issues}{solve these problems} would be implementing the distributed state estimation (DSE). In DSE, the power system will be divided into several smaller areas or sub-systems, and the SE process will \replaced[]{take place}{happen} \replaced[]{concurrently}{separately} in each area. A \replaced[]{low}{minimum} amount of information exchange at borders of the areas is required so that each area reaches convergence, i.e. the distributed network reaches to a similar solution as the centralized one. The amount of information that must be exchanged depends on the method applied. In \cite{asefi2020distributed}, a detailed comparison of the \replaced[]{recent}{current} DSE methods regarding indices such as convergence rate and information exchange has been made, which clearly confirms that each method varies from \added[]{one another considering} information transfer between areas.

The DSE algorithms can be classified into two categories, having a global control center, i.e. hierarchical DSE \cite{korres2010distributed, jiang2008diakoptic, zhao2005multi} or fully distributed \cite{minot2015distributed, marelli2015distributed, asefi2020distributed}. Both of them are successful in reaching to an acceptable solution compared to centralized algorithms. Alternating direction method of multipliers (ADMM) \cite{boyd2011distributed} that are in the category of distributed optimization \cite{conejo2007optimization}, have been very popular recently. In \cite{li2013robust} \deleted[]{applying the network gossiping method,} a fully decentralized adaptive SE scheme has been presented for the power system, \added[]{via applying the network gossiping method}. The method enables collaboration between areas to solve the global problem. However, there is still \replaced[]{a significant}{an insignificant} performance error in comparison to \replaced[]{CSE}{centralized SE} . Authors in \cite{xie2012fully} presented a \replaced[]{DSE}{fully distributed SE} for \replaced[]{wide area monitoring system}{WAMS}, which does not need local observability of all areas. In \cite{sharma2013multi}, a new multi-area SE method is discussed that utilizes a central coordinator; however, there is no need to exchange topology information between areas or from areas to the central coordinator. The proposed approach in \cite{guo2016hierarchical}, is a new hierarchical multi-area power system SE, which shares the sensitivity function of local estimators instead of boundary measurements or state estimates. As stated by the authors in \cite{guo2016hierarchical}, the approach reduces the information exchange, as well as increases convergence speed. In \cite{kekatos2012distributed}, the authors have provided an ADMM based DSE. Also, in \cite{minot2015fully} and \cite{minot2015distributed} a DSE process using matrix splitting method for DC and AC SE, respectively. For more details, we refer to \cite{gomez2011taxonomy} that presents a brief review of multi-area SE.

It is to be noted that \replaced[]{mostly}{usually} in the literature, the transmission system has been a matter of concern, which we \replaced[]{have followed}{try to follow} the same approach. \added[]{To solve AC SE via centralized method} or some of the distribution methods, such as matrix splitting or ADMM, would need linearization of the problem using Newton’s method. However, by applying the decomposition method \cite{conejo2007optimization} and the available solvers, there would be no further need for linearization of the problem.

%%%%%%%%%%%%%%%
\subsection{Blockchain}
% Traditional distributed systems have a centralised control system, where a single node controls and processes all the grid communications and the entities rely on this centralised architecture of communication. Centralised architecture is not a secure option as it has a single point of failure. If these central controlling nodes get compromised, all data can either get lost or controlled by attackers.

Blockchain (BC) is a peer-to-peer distributed ledger technology that stores data on multiple servers globally. In 2008, Satoshi Nakamoto’s whitepaper on Bitcoin \cite{Nakamoto2008} pioneered the use of BC technology in financial application\cite{Vigna2015}. \deleted[]{Initially,}BC technology was \added[]{primarily} used in the financial domain, \replaced[]{so as to \replaced[]{providing}{provide} trustfulness}{, i.e. trust} and secure environment without central authority where digital assets like cryptocurrency can \replaced[]{can be prevented from}{prevent} double-spending attacks\cite{Eyal2017,Peters2016}. Since then, the technology’s potential has moved beyond financial domains to different sectors like supply chain management, healthcare, etc. \cite{Madhwal2017,Malik2019a,Alzahrani2018,Korepanova2019,Mamoshina2017}.

BC is a distributed ledger of chronologically generated blocks containing cryptography linked blocks to the previous block forming a chain \deleted[]{, hence BC}. Any modification to the previous existing block will be reflected on every subsequent block, making it secure and immutable to modification. If an attacker changes data on any of the previous blocks, the following block’s data will also change, and the ledger can be compared with another \replaced[]{copy}{ledger} to track the point at which the data was manipulated and \added[]{later} rectified.  In cryptocurrency, it is computationally \replaced[]{hard}{impossible} to take control over the BC network because \replaced[]{the}{an} attacker will require $51\%$ of the network’s computing power, i.e., \replaced[]{it will be difficult for an attacker to fork from a past block and mine blocks faster, surpassing current (honest chain) block height. This will create double spending, which computationally hard}{preventing other miners from creating new blocks}\cite{Saad2019}. BC-based applications can provide security, trust, economic, and auditability\cite{ChoeLONDONCOIN:CRYPTOCURRENCY}. 

% \begin{figure}[h]
% \centerline{\includegraphics[width=\linewidth]{figures/BlockchainFigures/Centralised_Decentralised_Architecture.pdf}}
% \caption{Convergence curve of the state variables during the iteration for distributed method}
% \label{conv_WOD}
% \end{figure}

Since Bitcoin, many alternative cryptocurrencies (altcoins) have \replaced[]{emerged}{developed}. Ethereum\cite{Buterin2014} is the most popular cryptocurrency after Bitcoin, which \replaced[]{provides}{has provided} \replaced[]{an open-source platform to develop BC-based decentralized applications (DApps).}{a platform to develop BC projects on top of Ethereum BC. Ethereum is an open-source BC-based decentralized software platform.} \deleted[]{In 2015, Ethereum BC introduced a platform for BC smart contract development and decentralized applications (DApps), i.e.,} \added[]{DApps are application programs that runs on decentralized BC applications using Ethereum Virtual Machine (EVM)}. \deleted[]{DApps are digital applications running on a BC or peer-to-peer network of computers without controlling a single authority.} \added[]{For example,} smart contracts can specify the functionality and \deleted[]{conditions of working of the } \deleted[]{smart contract on the BC.} \deleted[]{For example, define a} condition, under which circumstances payment can occur between two individuals. These conditions are programmed and deployed on the BC, and individuals can abide by these conditions and transact in a secure environment without intermediaries. \replaced[]{EVM}{Ethereum} is one big computer that is made of small individual computers located globally. These computers are nodes connected, having a copy of the Ethereum BC. The transactions are broadcasted to the network via a node which is replicated across the network. 
For \replaced[]{feasibility demonstration}{Proof of Concept} of BC for secure data transmission, we have developed a prototype on the Ethereum platform using truffle framework \cite{Truffle2018} which can be deployed on local machines. We \added[]{have} created a smart contract, specified conditions and deployed it on a BC network running on local devices. \deleted[]{Sections \ref{sec:systemmodel} and \ref{sec:probForm} explaining the designed system architecture and problem formulation in detail and section \ref{sec:sim-res-dis} shows performance analysis.}

\added[]{Aside from financial applications of BC, it has been developed in other fields. For example, in \cite{qiu2019blockchain}, the authors propose a BC based method to preserve security of the spectrum sharing between aerial (unmanned aerial vehicle as a component of next generation cellular network) and terrestrial communication systems. Application of BC in the smart grid mainly has been investigated in the area of power markets, i.e. the issues related to the secure energy transactions \cite{musleh2019blockchain}. In \cite{aitzhan2016security}, a proof of concept (PoC) for decentralized energy trade using BC has been proposed, to enable peer to peer energy transactions. A BC based platform for solar energy trade amongst prosumers has been implemented in laboratory scale in \cite{pipattanasomporn2018blockchain}. However, a few works have been applied BC in power system for security purposes, \cite{liang2018distributed, dong2018blockchain}. These studies consider storing system wide measurement data in each measurement, which seems inefficient due to low memory of measurement units and time delay caused by encryption/decryption of the data.}

%%%%%%%%%%%%%%%
\replaced[]{Three main security features for the data in a smart grid are \textit{Confidentiality}, \textit{integrity} and \textit{availability}, which they refer to occasions when the data are accessible only to authorized users, the data are trustworthy in any operational circumstances, and the data are promptly and reliably available, respectively}{\textit{Confidentiality}, i.e. being accessible only to authorized users, \textit{integrity}, i.e. achieving the operational goals under harsh circumstances, and \textit{availability}, i.e. reliable availability of the information, are three main security features for a smart grid} \cite{mohan2020comprehensive}. Cyberattacks, such as the denial of service (DoS) or \added[]{false data injection (FDI)}, aim to \replaced[]{deteriorate}{destroy} \replaced[]{such}{these} properties. \deleted[]{As mentioned before,} Noting the case when a central control unit gets compromised, all data can either get lost or controlled by the attackers (\added[]{same case happened in Ukraine's cyber-attack} \cite{case2016analysis}), and \added[]{one of the potential} solutions could be a distributed control scheme. However, \deleted[]{based on the power system's communication structure,} the distributed grid can \added[]{also} be subjected to a cyber-attack, \replaced[]{i.e. attack to measurement units, to control centers, to communication line between control centers and measurement units, to communication line between control centers (i.e. between areas).}{while information transfer occurs between areas.} Due to the mentioned BC properties, in this paper \replaced[]{an integration}{a combination} of DSE with BC has been proposed to eliminate such an opportunity for the attackers \added[]{while information transfer occurs between areas.}

The work in this paper is inspired by \cite{kurt2019secure}. However, we have considered static SE, whereas they have considered dynamic SE. Dynamic SE refers to estimating the dynamic variables of the system (machine/dynamic load/distributed energy resources’ dynamic variables), and static SE (which already applicable in energy management system) hands out the algebraic variables of the system, i.e. voltage magnitude ($v$) and phase angle ($\theta$) for each bus \cite{zhao2020roles}. The second one is that in \cite{kurt2019secure}, the DC approximation of the power system has been considered while here the AC, SE has been studied. The third difference is considering the asynchronous behaviour of the information transfer within the power network that has not been provided in \cite{kurt2019secure}. And finally, detailed analysis and design of BC aided data transmission for DSE \deleted[]{that has not been addressed in the literature yet, due to the authors knowledge.} 

Beyond the numerous features of this research, the main contributions of this work can be summarized as follows:
\begin{itemize}
    \item Implementation of \added[]{BC based} DSE method, which requires \replaced[]{low}{minimum} data transfer and provides high accuracy.
    \item Application of \replaced[]{a}{the} new technology (i.e. BC) to increase the data transfer security in the power system. 
    \item \replaced[]{Consideration of asynchronous and delayed data transfer}{Realization of the  asynchronization and data transfer delays} that might happen within the \replaced[]{DSE}{distributed SE}.
    \item \replaced[]{Analysis and open-source code implementation for the design of BC aided data transmission in DSE.}{Detailed analysis and design of BC aided data transmission for DSE.}
\end{itemize}

The rest of the paper is organized as follows. In the second section, we present the modelling of SE, BC and \replaced[]{designed system architecture}{data transfer}. In the third section, the problem \replaced[]{formulation}{state} is presented. The fourth section provides the graphical and numerical \replaced[]{performance analysis}{results of the research}. Finally, in the section \ref{sec:conclusion} the research work is concluded.

%%%%%%%%%%%%%%%%%%%%%%%%%%%%%%%%%%%%%%%%%%
\section{System model}
\label{sec:systemmodel}

In this section, the mathematical equations governing the \added[]{DSE problem is presented. Decomposition method has been selected to solve DSE problem} \cite{conejo2007optimization}. \replaced[]{Later on, BC architecture and asynchronous data transfer are discussed.}{SE problem presented. First, we will provide the general problem formulation of SE. After that, the distributed perspective of the problem using the decomposition method is stated, and finally, the BC modelling for DSE has been provided.}

\subsection{Distributed state estimation}
\deleted[]{As mentioned before, }Given the noisy measurements \added[]{and network parameters} available in the power system, SE's role is to infer the state of the system. This study's measurements are considered as power flows, power injections \deleted[]{(along transmission lines)} and voltage magnitudes \deleted[]{(at the buses)}. In order to model the power network, considering the set of the measurements as $z$, the vector of errors associated with these measurements as $e$ and the set of nonlinear physical equations governing the power system that relates the state variables to the measurements as $f(x)$, we can state the following equation~\cite{gomez2018electric}: 
\begin{equation}
\begin{tabular}{c}
$z = f(x) + e$.
\end{tabular}
\label{eq:main_SE}
\end{equation}
It is noted that state variables, $x$, in this study are considered as voltage magnitudes and phase angles for each bus in the network\added[]{, taking into account the phase angle of slack bus (bus number one) as zero.} \deleted[]{Noting the redundancy of the network measurement units, SE's objective is to specify the most likely state values, having these units’ value.} Taking into account two assumptions for \deleted[]{the error corresponding to the measurements} $e$, that these errors are \added[]{mutually} independent and follow a normal distribution function, one can use the maximum likelihood method and \replaced[]{get}{reach} the following optimization problem:
\begin{equation}
\begin{tabular}{c}
    $\min \displaystyle\sum_{i=1}^{m}{W_i(z_i - f_{i}(x))^2}$
\end{tabular}
\label{eq:min-main-CE}
\end{equation}
where residual of $i^{th}$ measurement is defined as $r_i=z_i-f_i (x)$ and $W_i$ is the \replaced[]{weighting factor}{weight-related} to each measurement (inverse of the \added[]{squared} measurement variance, i.e. $W_i=\sigma_i^{-2}$). In DC SE, the~$f$~function would be approximated by a linear function \deleted[]{, and the system's final state would easily obtain}. However, in AC SE, we would need to use Newton’s method to linearize the problem and then solve it iteratively. \deleted[]{It is to be noted that the gain matrix's ill-conditioning might cause numerical problems cite{gomez2018electric, abur2004power}.}

% \subsection{Distributed state estimation}
\deleted[]{Smart grid overcome the mentioned problems in the previous section for centralized SE , e.g. policy and privacy or security/reliability, we can apply DSE.} \added[]{Without loosing generality, to apply DSE via decomposition method} the network would be divided into N areas \added[]{(or control centers)}, and the formulation represented in (\ref{eq:min-main-CE}) should be \replaced[]{separated}{broken down} for each area. Therefore the formulation of the optimization problem for each area can be represented as follows:

\begin{equation}
\begin{tabular}{c}
    $\min \displaystyle\sum_{i=1}^{m_N}{W_{N,i}(z_{N,i} - f_{N,i}(x_N))^2}$,
\end{tabular}
\label{eq:main-DSE}
\end{equation}
in which $z_{N,i}$ presents the measurements inside area $N$, $m_N$ is the number of measurements in area $N$, $W_{N,i}$ is the weighting factor corresponding to the measurement $z_{N,i}$ and $f_{N,i}$ is the equation related to this measurement. $x_N$ represents the set of state variables inside the \replaced[]{$N^{th}$}{$N$th} area, plus the set of auxiliary variables. The idea of auxiliary variables is to provide a consensus for the solution and it is due to the power line connections between areas. This will lead to the need for information \added[]{transfer} within the borders of these interconnected areas.
\deleted[]{In this work we have considered decomposition method to implement DSE on the test cases.} In section \ref{sec:probForm}, we will discuss the formulation of DSE and auxiliary variables in more details.

%SA: I think its better to introduce the abbreviation (BC) for blockchain and use BC afterwards
%SA: As requested by the format, figures should be refered to as, for example, Fig. 1
% not Figure 1.
%SA: No need to bold, Use italics for emphasis; do not underline. 

\subsection{Blockchain Architecture}
\begin{figure}[h]
    \centering
    \includegraphics[width=\linewidth]{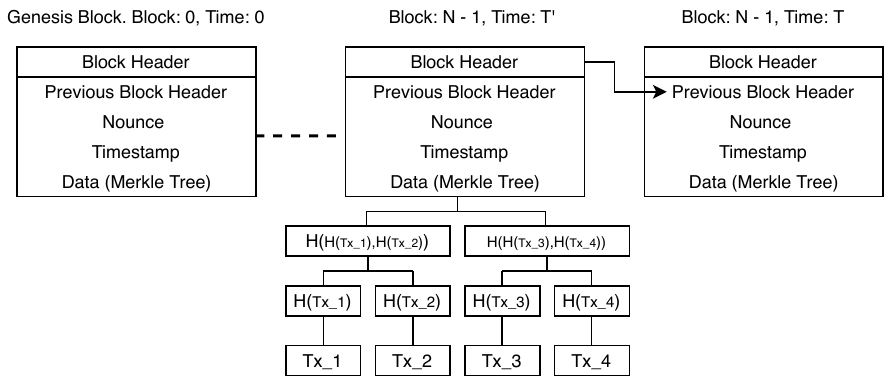}
    \caption{Data organization in blockchain}
    \label{Fig:BC}
\end{figure}

BC is a digital ledger of transactions distributed across the network of computer systems \deleted[]{on the BC}, with \replaced[]{atomic}{atomics} changes to the database\deleted[]{, i.e., transactions grouped into blocks}. The integrity and tamper-resistance of the transaction logs are assured because of the cryptographic hash linked among the blocks. BC is usually assumed to be decentralized architecture maintained by individual parties. Each node of the network owns a copy of the BC. Each BC block contains transactions, and every time a new transaction occurs on the BC, it is broadcasted to all nodes and added to a block along with other transactions waiting to get committed in a block.  This technology has developed over the last decade and can be categorised as private, public or consortium BC, each further divided by permissioned or permissionless. \added[]{As shown in Fig. \ref{Fig:BC},} every new block $N$ generated at time $T$ contains information from the previous $N-1$ block generated at time $T’$, where $T > T’$. \deleted[]{See Fig. ~\ref{Fig:BC}.}
 
\subsubsection{Consensus algorithm for decentralized ledger}
\added[]{}BC is a peer-to-peer network of nodes that function\added[]{s} individually without any central authority. Each node of the network can function individually, i.e. update ledger (creating and adding a block to the BC) and broadcast \replaced[]{new block}{it} to the other \replaced[]{nodes of the network}{networks} using the gossip protocol\cite{li2013robust}. The nodes verify the broadcasted block’s validity \deleted[]{in the network}, and \replaced[]{have to either}{the nodes agree to} accept or reject the proposed block, thus reaching a consensus. In distributed ledger technology, there exists a fundamental problem of reaching consensus. Majority of the BC projects \added[]{use} \replaced[]{any}{either} of the three most common consensus algorithms, i.e. \replaced[]{proof of work}{proof-of-work} (PoW), \replaced[]{proof of stake}{proof-of-stake} (PoS), and Byzantine \replaced[]{fault tolerant}{fault-tolerant}. \replaced[]{Similar to}{Like} Bitcoin, Ethereum uses a PoW consensus algorithm. In December 2020, Ethereum 2.0 was launched, which uses PoS consensus. In PoW BCs, block creators (\added[]{which are} called miners) are rewarded with mining rewards \replaced[]{along with}{and} transaction fees included in the block. This mining reward is the incentives for using computation power and electricity in finding the correct nonce within the target range. % In Bitcoin, after each $2016$ block, the target value is adjusted so that, on average, every $10$ minute a block is generated.

%$\texttt{Difficulty} = \texttt{DifficultyTarget} / \texttt{CurrentTarget}$. $\texttt{NextDifficulty} = \texttt{CurrentDifficulty} \cdot \frac{2 \; \texttt{weeks}}{T}$, where $T$ is the time taken to generate previous $2016$ blocks.

\deleted[]{The PoW consensus algorithm is used in generating blocks and \replaced[]{proposing}{proposed} to the other nodes of the network to add \added[]{them? it?} to the BC.}\replaced[]{Miners have}{Block proposing nodes has} to perform computation by running a hash of block’s content and incrementing a nonce until it produces a value less than the target. \added[]{Nounce is an integer that starts from 1 and increments until it produces a hash of block's content less than specific target value \cite{Nakamoto2008}}. Generating a hash on an arbitrary size input is a one-way function that produces a fixed output length\cite{Haber1991}, \added[]{ i.e., given input, we can generate an output of fixed-length, but not vice versa}. The hash function used is cryptographically secure and with brute force there exists a potential solution with complexity of \replaced[]{$O(2^n)$}{$O(2n)$} for $H(m) = H(m’)$, where $H$ is a SHA function\cite{SHA1} on an input $m$ and $m’$ and $m \neq m’$. This means that for a fixed output length on $n$, for example, $n = 256$ in the case of SHA256, the probability of success is \replaced[]{$k/2^{n}$}{2k-n}, where $k$ is a number of queries \cite{SHA256}.

\subsubsection{Ethereum Architecture}
A computer (node) can be a full node or light node \cite{ethereumnode} \replaced[]{running}{owning} an instance of the Ethereum BC. A full node stores the entire BC data and can serve any request. It verifies all blocks and states and can propose a new block to append on the ongoing chain. Light node stores only the header of the chain and can verify the validity of the state roots' data in a block header.  To interact with the DApp, clients should interact with the BC by running a full node by itself and using ethereum clients, like Geth, OpenEthereum, etc., to interact with the network. Ethereum BC has grown and consumes a significant storage amount and can be difficult to run a full node. Therefore, via a third-party platform like Infura, Alchemy, etc., \cite{infura, alchemy} provides application programming interface (API) to interact with ethereum BC feasible.

Ethereum comprises two main components:
\begin{itemize}
    \item 
        \textit{Database:} All activities on the network are recorded on the BC in the form of a transaction. Sending cryptocurrency from one address to another is recorded in a transaction with valid signatures and broadcasted to the network where other nodes commit to a block after verification. PoW consensus algorithms make \replaced[]{sure}{shires} that all the nodes in the network have the same BC data as all the valid transaction data. The data are stored in the form of a Merkle Patricia Tree. There are two types of addresses in Ethereum, Externally Owned Account (EOA), controlled by private keys and Contract Address, controlled by contract code. When a smart contract code is compiled and deployed from EOA, a contract address is created, and bytecode is stored in it. 
    \item 
        \textit{Code:} The smart contract is stored on the BC in a contract address in the form of code, known as byte code. \replaced[]{}{High-level programming languages like solidity cite{solidity} are compiled to low-level programming languages called Ethereum bytecode,  bytecode in human-readable form is known as opcodes.} The codes in contract addresses execute contract when a transaction is sent from EOA to \replaced[]{contract addresses}{Contract Addresses}.
\end{itemize}

For each transaction on the \added[]{Ethereum} BC, there is a fee known as Gas for executing transactions. Once a transaction is added to the block, the transaction fee goes to the miner as a reward for using computational resources. Gas is a unit to measure computation difficulty in \replaced[]{Ethereum Virtual Machine (EVM)}{EVM}. Gas is charged only when data are modified on the BC, i.e. reading and accessing data are not chargeable. Once the sender signs a transaction and broadcasts it, the Ethereum protocol debts gas fees in a fraction of ethers from the Ethereum account, lack of required gas amount will not allow the transaction to be \replaced[]{execute}{committed}. If there are no fees, attackers can flood the node’s memory pool with bogus transactions, causing \replaced[]{distributed DoS}{DDoS} attacks. Gas is not fixed for the transaction but \added[]{it is} variable \replaced[]{and depends}{depending} on the computational difficulty of a smart contract. 
%Smart contracts that can add two numbers will have less fee compared to a Decentralised Finance (DeFi) Application.
The sender of the transaction pays gas, and the miner who mines a block receives gas. Miner receives all the transaction gas that he includes in the block along with the block generation reward. Miners set the price of gas based on the computational power of the network required to process transactions and smart contract.

Since ether is not stable in value but sees daily change, therefore gas is a relative price converted to ethers based on the load on the network. In a congested network, the gas price will increase for each unit of gas. \deleted[]{So we should be able to switch from gas to ether.} So \replaced[]{There is }{we have} a gas price, i.e. how many units of ether are \replaced[]{transactor}{we} willing to pay for one gas unit.
% $\texttt{\texttt{gasCost}}(\texttt{ethere}) = \texttt{gasPrice} * \min(\texttt{gasCost}, \texttt{gasLimit})$.
Each opcode in Ethereum has a cost. The total cost of the contract is the summation of all the opcodes \cite{wood2014ethereum}. \replaced[]{}{$\texttt{gasLimit}$ is the limit of gas to spend in a transaction such that it does not drain all the amount for infinite loop transactions.}

The \replaced[]{EVM}{Ethereum Virtual Machine (EVM)} is a virtual stack embedded within each full Ethereum node that allows anyone to execute arbitrary bytecodes and plays a crucial role in the consensus engine of the Ethereum system. It allows anyone to execute arbitrary code in a \replaced[]{trustless}{trust-less} environment in which the outcome of execution can be guaranteed and is entirely deterministic. When you install and start the \replaced[]{Geth}{geth}, parity or any other client, the EVM is started, and it starts syncing, validating and executing transactions. The EVM is Turing complete, i.e. capable of performing any algorithm.

\begin{figure}[h]
    \centering
    \includegraphics[width=\linewidth]{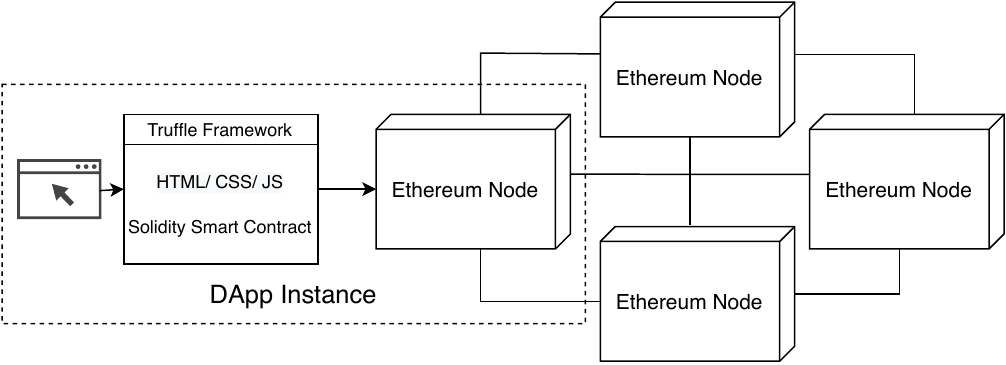}
    \caption{Ethereum network structure}
    \label{Fig:EVM}
\end{figure}

\subsubsection{Data Verification}
Before broadcasting the data that contains the formation of the transaction to other nodes in the network, the data should be signed using the private key. A signature is required to prove that the sender of the data are genuine and not an imposter who signed the message without the private key. BC uses asymmetric cryptography based on public key infrastructure. Like a physical signature, digital signatures are used to authenticate \added[]{electronically} a document\added[]{'s} contents like pdf, emails, etc. \deleted[]{electronically} \cite{ShenTu2015}. 

\deleted[]{Public-key or asymmetric cryptography is a cryptographic using keys, i.e. private keys or secret keys and public keys. The generation of key pairs depends on the one-way mathematical function.} \color{black} In the BC network, each node has its pair of public and private keys, and the public key is shared with all the other nodes. Owning a private key is equivalent to owning or controlling a node associated with its public key and accessing the activities restricted to it.

\begin{figure}[h]
    \centering
    \includegraphics[width=\linewidth]{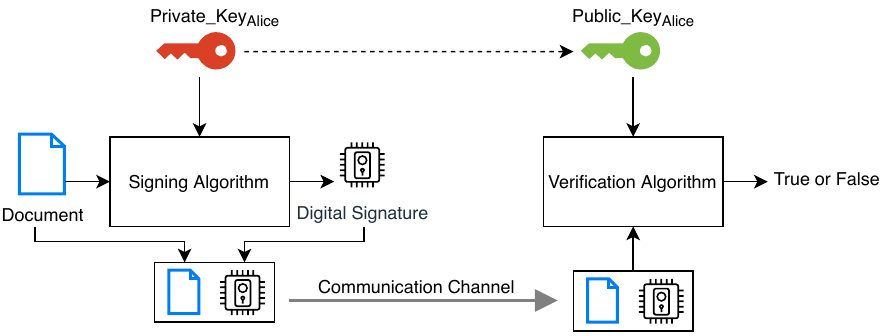}
    \caption{Data verification using private and public key}
    \label{Fig:DV}
    \vspace{-0.5 cm}
\end{figure}

To sign a message (or data), a function is calculated using the private key of the document’s sender. \replaced[]{The recipient’s using the public key of the sender, can verify if the document is correct and not tampered.}{At the recipient’s using public of the sender, we can verify if the document is correct and not tampered with.}

\subsection{Asynchronous data transfer}
\color{black}
The combination of renewable energy sources and information and communication technology (ICT) changes the power system's nature from a physical system to a CPPS~\cite{yohanandhan2020cyber}. Therefore, the physical part consists of a power grid, and the cyber part comprises a control and computation layer. \replaced[]{The physical layer consists of physical elements such as generators, transmission lines, transformers, etc. On the other hand, the cyber layer is responsible for computation, analysis and assessment of the power grid, and includes elements such as sensors, communication medium, control system, etc. \cite{yohanandhan2020cyber}.}{Although it is hard to provide a clear definition for the cyber layer due to the integration of power system elements with computational processors, we try to provide some features related to each part. The physical layer includes transmission lines, generators, transformers, etc., and physical laws govern it. The cyber layer is responsible for the computation, analysis and assessment of the power system on the regional and global scale.}
\replaced[]{A CPPS}{As a CPPS, the smart grid, or the power grid in general,} encounters different types of cyberattacks, such as \replaced[]{DoS and FDI}{the denial of service (DoS) or false data injection (FDI)}. However, latency attack has the potential to be considered as a new type of attack in the area of the power system, while it is already well-known for wireless network community \cite{chen2020system}. The power system undergoes a time delay of several milliseconds, while increasing this latency or time delay maliciously may lead to the power system instability \cite{chen2020system}.
Although the application of distributed methods and implementation of a BC based communication network may dissolve the issue \deleted[]{to a high extend}, still there would be a chance that the latency happens for the system. To study such a case, we have considered a delay in data transfer between areas in a randomized manner. In other words, at some iterations, an area randomly (based on uniform probability distribution) will be selected so as not to update its state variables. The comparison of the DSE results with and without delay is presented in section IV.

% Todo* Attack in GRID about changing data by eve and why we are combining it. :D *End Todo*
% *Todo* Problem statement for transfereing float data, i.e. we are using strings to write data on the blockchain *End Todo*

\section{Problem formulation}
\label{sec:probForm}
Suppose \added[]{that} we have divided the power system into $N$ areas, having $z_N$ measurements composed of power injection, power flow and voltage magnitude. Considering $x_k$ as the state variables related to area $k$ and $\Tilde{x}_l$ as the auxiliary variables estimated by area $k$ related to its neighboring area $l$, one can rewrite (\ref{eq:main-DSE}) into the following equation:
\begin{equation}
     \displaystyle\min_{x_k}\text{ } f_k(x_k)+\sum_{l\in\Lambda_k}{f_{kl}(x_k,\Tilde{x}_{l})},
\label{eq:decomposition-main-DSE}
\end{equation}
where $\Lambda_k$ indicates the set of all neighboring areas of $k^{th}$ area. 
It is clear that (\ref{eq:decomposition-main-DSE}) is composed of two statements. The first statement is related to the measurements that the physical equation for calculating them only requires the state variables inside the area and can be written as follows:
\begin{eqnarray}
     & \added[]{\scriptstyle f_k(x_k)=\sum_{i\in\Lambda_k^v}{W_{k,i}^v(v_{k,i}^{m}-v_{k,i})^2}} \nonumber \\ [0.2cm] 
    & {\scriptstyle + \sum\limits_{i\in\Lambda_k^P}{W_{k,i}^P(P_{k,i}^{m}-P_{k,i}(.))^2}+\sum\limits_{i\in\Lambda_k^Q}{W_{k,i}^Q(Q_{k,i}^{m}-Q_{k,i}(.))^2}} \\
    \nonumber &
   {\scriptscriptstyle + \hspace{-0.1cm} \sum\limits_{(i,j)\in\Lambda_k^{PF}}{W_{k,ij}^{PF}(P_{k,ij}^{m}-P_{k,ij}(.))^2}
   +\hspace{-0.1cm}\sum\limits_{(i,j)\in\Lambda_k^{QF}}{W_{k,ij}^{QF}(Q_{k,ij}^{m}-Q_{k,ij}(.))^2},}
    \label{eq:decomposition-DSE-firstpart}
\end{eqnarray}
\replaced[]{where $i$ and $j$ are arbitrary buses; $\Lambda^{v}_{k}$, $\Lambda^{P}_{k}$, $\Lambda^{Q}_{k}$, $\Lambda^{PF}_{k}$ and $\Lambda^{QF}_{k}$ indicate the set of voltage, active power injection, reactive power injection, active power flow and reactive power flow measurements in area $k$, respectively; $W_{(.)}^{(.)}$ weighting factor for the measurements; $P^m_{(.)}$, $Q^m_{(.)}$ and $v^m_{(.)}$ are the active power injection or power flow, reactive power injection or power flow and voltage observed measurements, respectively; While $P_{(.)}$, $Q_{(.)}$ and $v_{(.)}$ are the physical equations of these measurements. These physical equations governing the power system are provided in appendix.}{where $\Lambda_{(.)}^{(.)}$ indicates the set of measurements; $W_{(.)}^{(.)}$ weighting factor for the measurements; $P^m_{(.)}$, $Q^m_{(.)}$ and $v^m_{(.)}$ are the active power injection or power flow, reactive power injection or power flow and voltage measurements; while $P_{(.)}$, $Q_{(.)}$ and $v_{(.)}$ are the physical equations of these measurements. These physical equations governing the power system are provided in appendix. }

The second statement of (\ref{eq:decomposition-main-DSE}), is related to the measurements in $k$ that need to receive state values regarding the buses in connection with the neighboring area $l$. It is to be noted that, for calculation of the physical equations regarding these measurements, we use the auxiliary variables:
\begin{eqnarray}
\label{eq:decomposition-DSE-secondpart}
    &\added[]{\scriptstyle f_{kl}(x_k,\Tilde{x}_{l})= \sum_{i \in \Lambda_{kl}^P}{W_{kl,i}^P(P_{kl,i}^{m}-P_{kl,i}(.))^2}} \nonumber \\ [0.2cm]
    &{\scriptstyle +\hspace{-0.1cm}\sum\limits_{i \in \Lambda_{kl}^Q}{W_{kl,i}^Q(Q_{kl,i}^{m}-Q_{kl,i}(.))^2}+\hspace{-0.1cm}\sum\limits_{(i,j)\in\Lambda_{kl}^{PF}}{W_{kl,ij}^{PF}(P_{kl,ij}^{m}-P_{kl,ij}(.))^2}} \nonumber \\
    &{\scriptstyle + \hspace{-0.1cm}\sum\limits_{(i,j)\in\Lambda_{kl}^{QF}}{W_{kl,ij}^{QF}(Q_{kl,ij}^{m}-Q_{kl,ij}(.))^2}+ \hspace{-0.1cm}\sum\limits_{i\in\Lambda_{kl}}{W_{k,i}^{\Tilde{v}}(v_{l,i}-\Tilde{v}_{l,i})^2}} \\ \nonumber 
    &{\scriptstyle + \hspace{-0.1cm}\sum\limits_{i\in\Lambda_{kl}}{W_{k,i}^{\Tilde{\theta}}(\theta_{l,i} - \Tilde{\theta}_{l,i})^2}}, 
\end{eqnarray}
 where $\Tilde{\theta}_{(.)}$ and $\Tilde{v}_{(.)}$ are the auxiliary variables. It is worth noting that the last two statements in (\ref{eq:decomposition-DSE-secondpart}) are utilized to provide a consensus for this minimization function.

\subsection{Proposed Blockchain Solution}
Building \replaced[]{DSE}{grid}’s data transmission architecture \replaced[]{based on}{on top of} BC provides a security feature of the technology to transfer data among \replaced[]{system}{grid} areas. BC integration can ensure honesty in the system as the transaction's sender can only sign each transaction. \replaced[]{}{Thus increasing trust and security to resolve any dispute can arise for an incorrect transaction, which is less likely.}

A PoC is developed on the Ethereum test network and deployed using Truffle framework and Ganache. Ethereum provides tools to build smart contracts and decentralized applications without any downtime or any third-party interference. Truffle Suite is a  BCs development environment, testing framework, and asset pipeline using the \replaced[]{EVM}{Ethereum Virtual Machine (EVM)}. Ganache \cite{Ganache} is a personal BC for Ethereum and Corda based distributed application development. \replaced[]{Utilizing}{With} Ganache and Truffle, the entire DApp can be developed in a safe and deterministic environment. \deleted[]{Ganache UI is a desktop application supporting both Ethereum and Corda technology.} \added[]{The code repository containing open source prototype is available in \cite{yashGridBC2020}.}

The \replaced[]{EVM}{Ethereum virtual machine} has separate storage areas: 
\begin{itemize}
\item All contracts have state variables, and the state variables are stored on the BC, i.e. \replaced[]{the}{The} data are recorded into the BC itself. When the contract executes some code, it can access all the previously stored data \replaced[]{in}{into} its storage area. \replaced[]{}{This storage is quite expensive.}

\item Memory holds temporary values and only exists in the calling function and has less gas price because the stored memory gets erased between calls. Gas price increases with the size of memory scaling quadratically. Though, comparatively cheaper than storage.

\item The stack holds small local variables, and here the computations happen. This data can only hold a limited amount of values up to 1024 small local variables. 

\item \replaced[]{}{Logs store data but may not be considered memory type as it is not written on a transaction but blocks.} Logs store data in an indexed structure with mapping, and with filters, specific data can be accessed. Logs are inaccessible to contract but are mainly used for events that occur on the BC.
\end{itemize}

\subsection{System Overview}
The proposed BC solution focuses on establishing a secure architecture of transferring arbitrary data for every iteration among the \replaced[]{DSE areas}{grid entities} based on the established connections on the BC.  Fig. \ref{Fig:IEEE14SysTop} shows the main participating entities of the system:

\begin{itemize}
    \item \textit{\replaced[]{DSE areas}{Grid area}}: The control center at each area is responsible for receiving \added[]{data and then, calculate} SE and after that send\deleted[]{ing} data to another area. 
    \item \textit{Auditor}: Provides public key infrastructure \cite{pki1996} \added[]{to all DSE areas} and is responsible for maintaining smart contracts on the BC and can establish or demolish connection between two areas. \added[]{In other words, only the auditor can establish communication between two or more than two areas by sending a transaction to the smart contract address that sets communication to \textit{true} between areas on the smart contract. Auditor is like a supervising body of the infrastructure of the DSE network. Although, it is responsible for deploying contracts on the blockchain, the DSE areas can communicate, i.e. transfer date, with each other via smart contract without interference from the auditor.} If any issues arise on the BC, the auditor can resolve this issue with the BC. The \replaced[]{DSE data}{grid} transactions are independent, and the auditor is not involved.

\end{itemize}

\deleted[]{Each of the entities has its Ethereum address and can interact with the smart contracts created deployed by the auditor. Although the auditor is the owner of the smart contract, the data transfer in grid happens independently.}

\subsection{System Design}
On the BC, two contracts are deployed. First, to establish/demolish connection between the \deleted[]{grid} areas. Second, to transfer data per iteration between the  \deleted[]{grid} areas \replaced[]{within}{with} the established connection. The following section describes the details.

\subsubsection{Establishing/Demolishing Connection}
\replaced[]{The auditor manages the connections between areas through algorithm \ref{op0}. The smart contract \replaced[]{emit}{throws} event upon each connection change to inform all the areas.}{The auditor controls algorithm \ref{op0} call, and only he is allowed to establish and demolish connection between two grids area. For each new connection, the network emits events of a new connection to all the area.} 

\subsubsection{Data Transfer}
Algorithm \ref{op1} smart contracts listens to all the transaction call of the first deployed smart contract and as per update the state of the connections of this smart contract. This algorithm takes four parameters i.e. \textit{sender}, \textit{receiver}, \textit{iteration} and \textit{payload}. Each \deleted[]{grid} area in our case study has a different data payload size \deleted[]{, either two or four} (i.e. state variables which needs to be transferred). With each iteration, data are passed as arrays of float integers as string type because it is impossible to pass a negative number in a smart contract. With each transaction of the iteration, the transaction event is emitted and notified to the receiving \replaced[]{area}{grid}, who can process the data off-chain as peruse. 

\deleted[]{I suggest to remove all underlines from the algorithms}

% First Algorithm
\begin{algorithm}[h]
\caption{Establish/Demolish area Connection}
\begin{algorithmic}[1]
\renewcommand{\algorithmicrequire}{\textbf{Input:}}\label{op0}
\REQUIRE address\_deployer, address\_from, address\_to
% \ENSURE 
\\\textit{Initialization:} $connection(from,to) \leftarrow bool$
% \STATE first statement
% \\ \textit{LOOP Process}
% \FOR {$i = l-2$ to $0$}
% \STATE statements..
\IF {($msg.sender \ne address\_deployer$)}
\STATE from $\leftarrow$ address from
\STATE to $\leftarrow$ address to
\IF{($from \ne to$)}
\IF{($connection(from,to) \ne True$)}
\STATE Set connection(from,to) $\leftarrow$ True
\ELSE
\STATE Revert and show error "Connection exist"
\ENDIF
\ELSE
\STATE Revert and show error "No Self Connection"
\ENDIF
\ELSE
\STATE Revert and show error "Only Owner Access"
\ENDIF
% \ENDFOR
\end{algorithmic}
\end{algorithm}
% Second Algorithm 

\begin{algorithm}[h]
\caption{Data Transfer}
\begin{algorithmic}[1]
\renewcommand{\algorithmicrequire}{\textbf{Import:}}\label{op1}
\renewcommand{\algorithmicensure}{\textbf{Input:}}
\REQUIRE ‘Establishing Demolishing Connections’
\ENSURE address\_sender, address\_to, interation\_number, data\_String
% \\\textit{Initialisation:} $d$
% \STATE first statement
% \\ \textit{LOOP Process}
% \FOR {$i = l-2$ to $0$}
% \STATE statements..
\IF {($msg.sender \ne address\_sender$)}
\IF{($connection(from,to) = True$)}
\STATE Call function to transact these values on blockchain
\STATE Notify transaction in the network
\STATE Apply the transferred data in the current iteration for state estimation
\ELSE
\STATE Revert and show error "No Connection"
\ENDIF
\ELSE
\STATE Revert and show error "Only msg.senders"
\ENDIF
% \ENDFOR
\end{algorithmic}
\end{algorithm}
\subsection{Security}
This architecture provides security because each transaction requires a transaction signature. The connection can only be established by the smart contract owner as there is a specific check-in of the smart contract that requires signature verification. Signature is created using the private key, and address generation also requires a private key. Therefore, losing the private key, especially by the auditor, i.e. controller of the architecture, can compromise the whole system. 

\section{simulation results and discussion}
\label{sec:sim-res-dis}
\begin{figure}[h!]
\centerline{\includegraphics[width=\linewidth, height=160 pt]{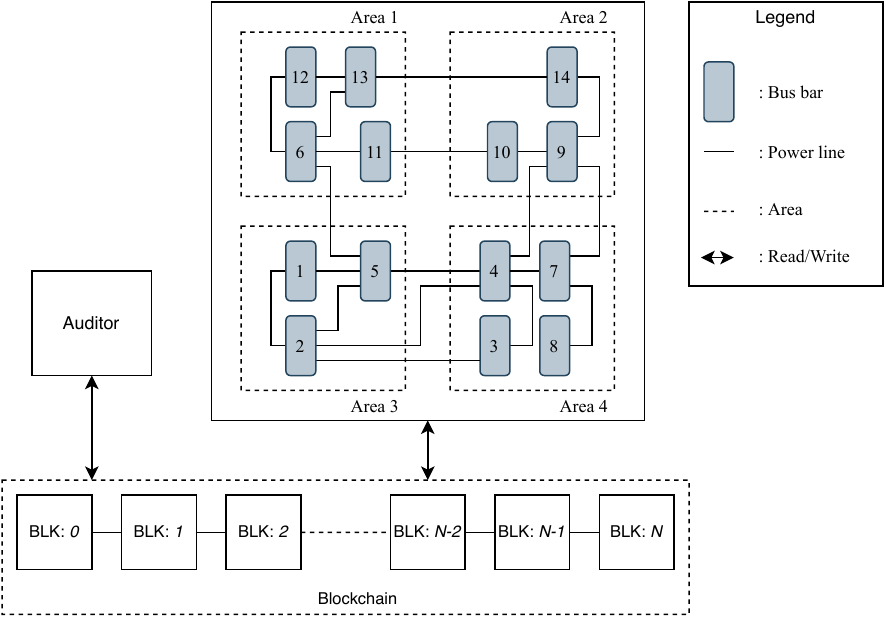}}
\caption{Distributed topology of the IEEE 14 bus system \cite{minot2015fully} integrated with blockchain}
\label{Fig:IEEE14SysTop}
\vspace{-0.4 cm}
\end{figure}

In this section, \replaced[]{the test case (i.e. IEEE 14 bus system \cite{christiepower}) results are presente utilizing the proposed method}{the proposed method's results on a test case, i.e. IEEE 14 bus system, are presented}. The system has been divided into four areas \replaced[]{and}{.} Fig. \ref{Fig:IEEE14SysTop} shows the topology of the studied test case.

\replaced[]{In this research, AC SE has been considered}{However, we have considered static AC SE}, where the state variables would be voltage magnitudes and phase angles at each bus. \deleted[]{It is to be noted that the measurements are consist of voltage magnitude, power flows and injections (both active and reactive)}. \added[]{The number of state variables and measurements are $27$ and $41$, respectively.}
The \replaced[]{weighting factor for all measurement units has been considered equal to $10^{4}$}{noise covariance for all measurement units has been considered $10^{-4}$}. \added[]{In order to solve \eqref{eq:decomposition-main-DSE}, MATLAB (version \textit{R2018b}) solver (Sequential quadratic programming) has been applied and for initiation of the optimization process} the initial value for state variables \replaced[]{have been set to}{are a} flat start \deleted[]{situation}, i.e. voltage magnitude of “1” and phase angle value of “0”. Moreover, the bus number one has been selected as the slack bus \added[]{with phase angle zero}. 
To evaluate the prototype's performance, the smart contract was deployed on a local BC server and interacted with the python application. The experiments were performed on a computer with memory 16 GB 2400 MHz DDR4, Intel Core i9 running @2,3GHz. 
% I remove this because you have mentioned the system, let it be like that.
%% We have implemented the simulation in MATLAB R2018b on a computer with Intel(R) Core i5 processor and 8 GB of RAM.

As mentioned before, we have considered two different cases. \replaced[]{The data transfer between areas are simultaneously in the first case and with latency (time delay) in the second case.}{The first case is where the data transfer between areas are considered simultaneously. The next one is when we consider the latency or time delay.} The graphical and numerical results of both cases are presented here.
\vspace{-0.3 cm}
\begin{figure}[h!]
\centerline{\includegraphics[width=\linewidth,height=160 pt]{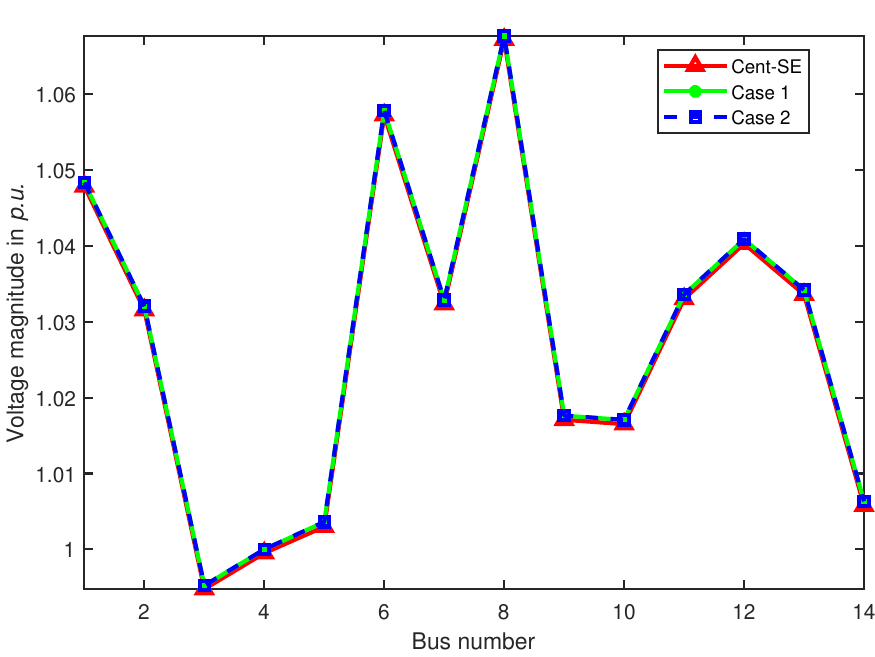}}
\caption{IEEE 14 bus system voltage magnitude for centralized (Cent-SE) and distributed state estimation interacting with blockchain (case 1 and case 2)}
\label{Fig:voltage_WOD}
\vspace{-0.2 cm}
\end{figure}

\begin{figure}[h!]
\centerline{\includegraphics[width=\columnwidth,height=160 pt]{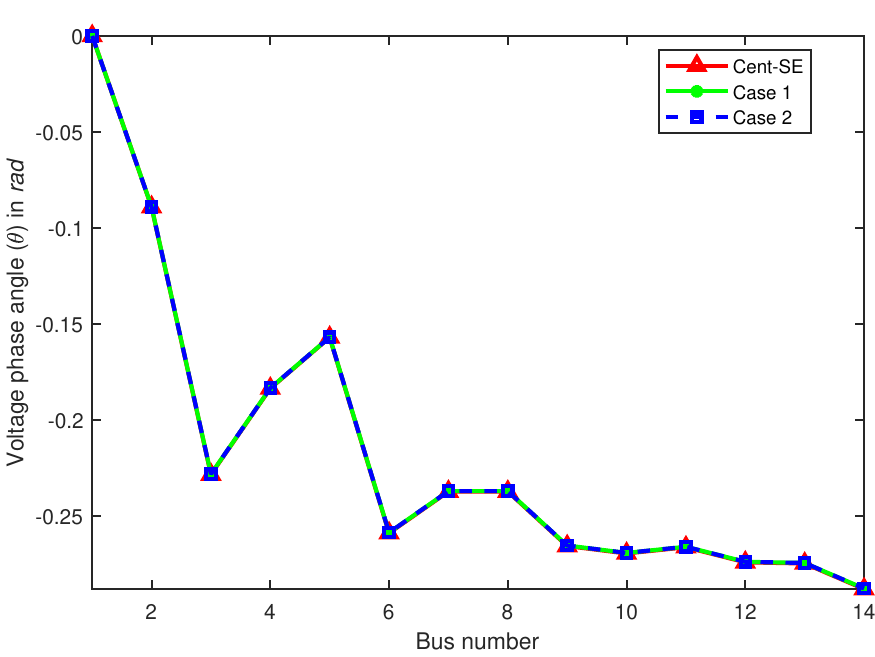}}
\caption{IEEE 14 bus system voltage phase angle for centralized (Cent-SE) and distributed state estimation interacting with blockchain (case 1 and case 2)}
\label{Fig:phase_WOD}
\vspace{-0.4 cm}
\end{figure}

Fig. \ref{Fig:voltage_WOD} and Fig. \ref{Fig:phase_WOD} represent the comparison of centralized and distributed estimated voltage magnitude and voltage phase angle for IEEE 14 bus test system interacting with BC. The distributed method has succeeded to reach the centralized values in both cases.

\begin{figure}[h!]
\centerline{\includegraphics[width=\columnwidth,height=160 pt]{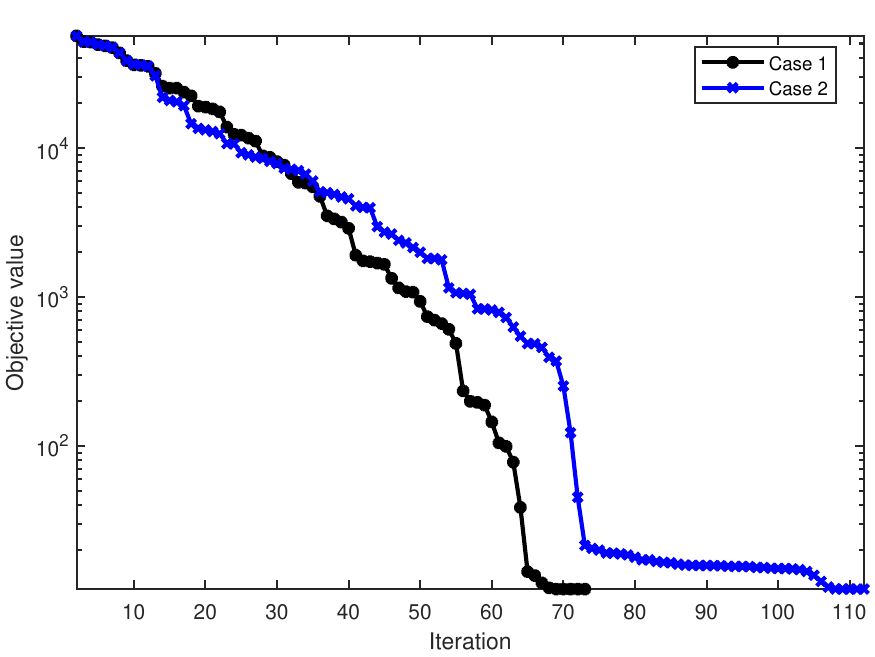}}
\caption{Distributed method objective value during iteration for case 1 and case 2}
\label{Fig:obj_WOD}
\vspace{-0.2 cm}
\end{figure}

Fig. \ref{Fig:obj_WOD} shows the distributed method objective value during the IEEE 14 bus system's optimization procedure. As proposed in \cite{asefi2020distributed}, we have considered the state variables convergence rate as convergence criterion. \added[]{It means that the difference between obtained state variables of two successive iterations are measured at each area and if the value is below the specified threshold (it has been set to $10^{-6}$ \cite{asefi2020distributed}), the optimization stops.} It is clear that in case 2, where there is a delay in data transmission, the number of optimization iteration increases.

% \begin{figure}[h!]
% \centerline{\includegraphics[width=\linewidth]{figures/DSEfigures/conv_WOD.eps}}
% \caption{Convergence curve of the state variables during the iteration for distributed method}
% \label{Fig:conv_WOD}
% \end{figure}

% \begin{figure}[h!]
% \centerline{\includegraphics[width=200]{figures/DSEfigures/residu_WOD.eps}}
% \caption{Comparison of the residuals}
% \label{Fig:residu_WOD}
% \end{figure}

\deleted[]{Finally,} The numerical results of the comparison between CSE and DSE are presented in table \ref{tab:final-res-DSE}. The iteration number and objective value of both centralized and distributed are presented\deleted[]{, respectively}. As mentioned in section \ref{sec:systemmodel}, the objective value for \replaced[]{CSE}{centralized SE} is obtained using~(\ref{eq:main_SE}) and applying Newton's method. However, for DSE, after solving the optimization problem stated in (\ref{eq:decomposition-main-DSE}) for all areas, we gathered the state variables and placed these state variables into (\ref{eq:main_SE}). The objective\replaced[]{ values are obtained}{'s obtained result is} by \replaced[]{substituting}{feeding} the DSE state variables into (\ref{eq:main_SE}).

\begin{figure}[h!]
\centerline{\includegraphics[width=\linewidth, height=180 pt]{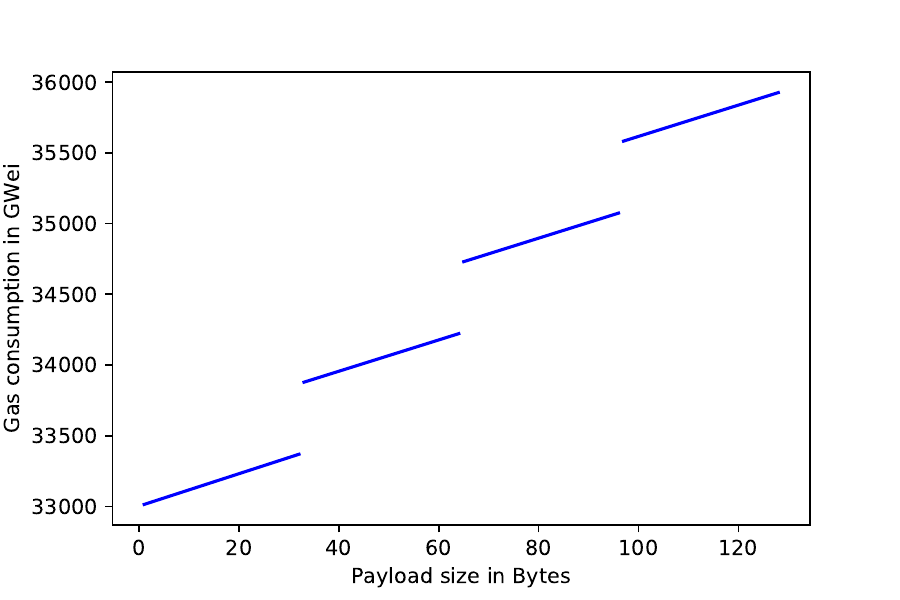}}
\caption{Gas consumption in Gwei to transfer bytes with payload size}
\label{Fig:result2}
\vspace{-0.5 cm}
\end{figure}

\begin{figure}[h!]
\centerline{\includegraphics[width=\linewidth, height=190 pt]{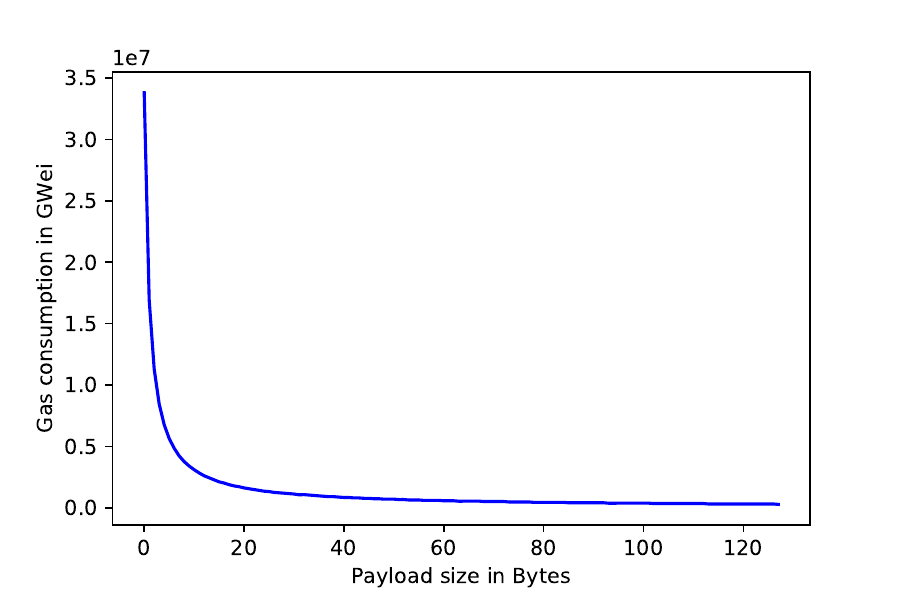}}
\caption{Gas consumption to transfer 1024 bytes per payload size in Bytes}
\label{Fig:result3}
\vspace{-0.3 cm}
\end{figure}

\begin{table}[h!]
\centering
\caption{Numerical results of comparing centralized and distributed method for case 1 and case 2}
\label{tab:final-res-DSE}
\resizebox{\columnwidth}{!}{%
\begin{tabular}{ccccccc}                                 & \multicolumn{2}{c}{\textbf{Iteration}} & \multicolumn{2}{c}{\textbf{Objective}} & \textbf{Objective } &  \textbf{Distributed }\\
\multirow{-2}{*} &      CSE         &       DSE        &        CSE       &             DSE   & \textbf{ error} & \textbf{ objective} \\
\textbf{case 1}                                                 &     6          &      73         &              11.5568 &     11.6801          & 1.0559 \% & 10.9951 \\
\textbf{case 2}                                                 &       6        &      112         &               11.5568 &       11.6804        & 1.0582 \% & 10.9951
\end{tabular}%
}
\end{table}
As shown in the table, the factual error between these values is approximately 1 per cent \deleted[]{for significantly reliable cases}. \replaced[]{The necessity of considering objective value is due to the fact that}{It is essential to consider objective value between the centralized and distributed method because} \color{black} one of the \deleted[]{standard} methods to specify measurement anomalies, so-called \textit{bad data}, is to \deleted[]{identify maximum residual and} compare \replaced[]{objective value}{it} with the chi-square value \cite{gomez2018electric}. So, considering measurement residuals distributed method matches the centralized to a great extent as well.

\added[]{The last column of table \ref{tab:final-res-DSE} is devoted to the DSE objective value, which is slightly different from the centralized objective value. This slight difference is quite apparent, and it is due to the application of the auxiliary variables. In other words, applying auxiliary variables in (\ref{eq:decomposition-DSE-secondpart}) is the main reason leading to a better objective value. There might be a method to utilize the benefit of applying these auxiliary values to improve SE results further, but it is beyond the scope of the paper.}

Fig. \ref{Fig:result2} shows the result of the experiment to check the gas consumption, \added[]{amoount of gas used to execute a transaction,} with respect to the transaction payload size in bytes of different values in a transaction, i.e. in a hexadecimal value and used to check how it will influence the processing time. \replaced[]{Different value precision results in different payload sizes. We executed $128$ transactions of payload size one and bytes of size $k$ from $1$ to $128$.  In EVM, a one-word is a maximum of $32$ bytes. Zero bytes pad each payload up to the closest factor of $32$ bytes and processed as a sequence of $32$ bytes words. Most of the operation consumption goes to cryptographic signature checks by the nodes. Gas consumption varies with different byte sizes, and we can see a significant shift for each consecutive $32$ bytes, but within each set, the gas fees increased linearly with an increment of a byte.}{It is assumed that for each transaction between areas, the payload is of size one. We assume set of $4$ payload size $2^{k-1}$ in character where $k  \in  \{1, \dots, 6\}$, i.e. half the length of bytes. Gas consumption varies with different payload size, and we can see that it is somehow linear with quite a significant shift but a small slope. The majority of the operation consumption goes to cryptographic signature checks by the nodes. Hence conclude that the processing time will depend on the number of the transaction.}

Fig. \ref{Fig:result3} indicates the optimization of the transfer procedure where several transactions can be concatenated \replaced[]{as}{into} one string, i.e. bulk data transfer\replaced[]{. This would result in}{, thus} less \added[]{number of} transaction to transfer the same amount data without spending extra gas \added[]{for each execution}. \replaced[]{For the experiment, we measured the gas consumption to transfer 1024 bytes per $2^{k-1}$ bytes where $k  \in  \{1, \dots, 8\}$, with increase on payload size, the gas consumption reduces for computation at nodes.} {Each  transaction  will  increase  the  payload  size  and  reduce gas consumption for computation at nodes.}

\deleted[]{The last column of table \ref{tab:final-res-DSE} is devoted to the DSE objective value, which is slightly different from the centralized objective value. This slight difference is quite apparent, and it is due to the application of the auxiliary variables. In other words, applying auxiliary variables in (\ref{eq:decomposition-DSE-secondpart}) is the main reason leading to a better objective value. There might be a method to utilize the benefit of applying these auxiliary values to improve SE results further, but it is beyond the scope of the paper.}

\section{Conclusion}
\label{sec:conclusion}
Blockchain technology has attracted research and industrial communities' attention due to its diverse and novel characteristics. Needless to say that the future power grids, so-called smart grids, can benefit from these features in different industrial divisions. In this regard, we tried to point out blockchain application in smart grids' main sector\deleted[]{s}, \replaced[]{i.e. the state estimator}{the energy management system}. State estimation plays a vital role in regulating system operator decisions such as \replaced[]{contingency analysis,}{the} electricity market and load forecasting in energy system management.

In this work, we have proposed a combination of distributed state estimation and a blockchain designed communication platform for secure data transmission and increasing the system's reliability. Application of the smart contract concept would lead to improving the security of the overall system. Moreover, the robustness of the method against the data transmission latency has been analysed.

As mentioned before, we introduced a scheme for the combination of state estimation with blockchain in a distributed transmission system. Therefore, implementing such a combination for the distribution system, in which the applications of renewable energy sources are increasing exponentially, can be a future direction. Another research direction for the future can be introducing multi-signature that will make this architecture more secure. Additionally, economical analysis for BC's implementation in the power system would be of interest to research and the industrial community and can be considered \added[]{as} another future direction for this study. 

\section*{Appendix}

\vspace{-0.1 cm}
Measurement units that have been considered in this study are composed of voltage measurement, real power injection, reactive power injection, real power flow, reactive power flow. In this section physical equations governing the power system are provided.
\vspace{-0.1 cm}
\subsection{Active and reactive power injection and power flow (inside area k)}
\vspace{-0.2 cm}
\begin{footnotesize}
\begin{eqnarray}
    & P_{k,i}(.) = \displaystyle\sum_{i=1}^{n}{v_{k,i}v_{k,j}(G_{ij}\cos{\theta_{k,ij}} + B_{ij}\sin{\theta_{k,ij}}}), \nonumber \\
    & Q_{k,i}(.) = \displaystyle\sum_{i=1}^{n}{v_{k,i}v_{k,j}(G_{ij}\sin{\theta_{k,ij}} - B_{ij}\cos{\theta_{k,ij}}}), \nonumber \\
    & P_{k,ij}(.) = v_{k,i}v_{k,j}(G_{ij}\cos{\theta_{k,ij}} + B_{ij}\sin{\theta_{k,ij}})-\text{ }G_{ij}v^2_{k,i}, \nonumber\\ [0.2cm] 
    & Q_{k,ij}(.) = v_{k,i}v_{k,j}(G_{ij}\sin{\theta_{k,ij}} - B_{ij}\cos{\theta_{k,ij}}) \nonumber  \\
    & +v^2_{k,i}\left(B_{ij} - \frac{b_{ij}^s}{2}\right), \nonumber 
\label{eq:physical-pf-pi}
\end{eqnarray}
\end{footnotesize}
where $G_{ij}$ and $B_{ij}$ is the real and imaginary part of the element in the \replaced[]{$i^{th}$}{$i$th} row and \replaced[]{$j^{th}$}{$j$th} column of the network admittance matrix, respectively; And $\frac{b_{ij}^s}{2}$ is the shunt suseptance considering the $\pi$ equivalent mode of the line.
\vspace{-0.4cm}
\subsection{Active and reactive power injection and power flow (between area k and area l)}
The equations between areas would be the same as inside areas, but the only difference would be the value of state variables. If the state variable is for the neighboring area we should use the auxiliary variables. For example, if the \replaced[]{$j^{th}$}{$j$th} bus is for the neighboring area, we should use $\Tilde{v}$ and $\Tilde{\theta}$.  

%%%%%%%%%%%%%%%%%%%%%%%%%%%%%%%%
\bibliographystyle{IEEEtran}
\bibliography{generic-color.bib}

\end{document}